\documentclass[letterpaper, 10 pt, conference]{IEEEtran}

\usepackage[final]{changes}
\usepackage{amsmath, amssymb,amsfonts}
\usepackage{graphicx}
\usepackage{multicol}
\usepackage{tabularx}
\usepackage{subfig}
\usepackage{soul}
\usepackage{float}
\usepackage{multirow}
\usepackage[ruled,longend]{algorithm2e}
\usepackage[backend=biber, sorting=none]{biblatex}
\usepackage{siunitx}
\usepackage{color,xcolor,comment}
\newcolumntype{P}[1]{>{\centering\arraybackslash}p{#1}}

\IEEEoverridecommandlockouts

\title{\LARGE \bf
Using Physiological Information to Classify Task Difficulty in Human-Swarm Interaction
}
\author{Joseph P. Distefano$^1$, Hemanth Manjunatha$^1$, Souma Chowdhury$^1$,\\ Karthik Dantu$^2$, David Doermann$^2$, and Ehsan T. Esfahani$^{1*}$% <-this % stops a space

\thanks{This work was supported by National Science Foundation Award IIS-1927462 and the DARPA Cooperative Agreement HR00111920030.}
\thanks{$^{1}$ JD, HM, SC, and EE are with the Department of Mechanical and Aerospace Engineering, University at Buffalo, Buffalo,
NY, 14260 USA.}%
\thanks{$^{2}$ KD and DD are with the Department of Computer Science and Engineering, University at Buffalo, Buffalo,
NY, 14260 USA.}%
\thanks{$^*$Corresponding Author, 
{\tt\small ehsanesf@buffalo.edu}}
}

\begin{document}

\maketitle
\thispagestyle{empty}
\pagestyle{empty}

%%%%%%%%%%%%%%%%%%%%%%%%%%%%%%%%%%%%%%%%%%%%%%%%%%%%%%%%%%%%%%%%%%%%%%%%%%%%%%%%
\begin{abstract}

Human-swarm interaction has recently gained attention due to its plethora of new applications in disaster relief, surveillance, rescue, and exploration. However, if the task difficulty increases, the performance of the human operator decreases, thereby decreasing the overall efficacy of the human-swarm team. Thus, it is critical to identify the task difficulty and adaptively allocate the task to the human operator to maintain optimal performance. In this direction, we study the classification of task difficulty in a human-swarm interaction experiment performing a target search mission. The human may control platoons of unmanned aerial vehicles (UAVs) and unmanned ground vehicles (UGVs) to search a partially observable environment during the target search mission. The mission complexity is increased by introducing adversarial teams that humans may only see when the environment is explored. While the human is completing the mission, their brain activity is recorded using an electroencephalogram (EEG), which is used to classify the task difficulty. We have used two different approaches for classification: A feature-based approach using coherence values as input and a deep learning-based approach using raw EEG as input. Both approaches can classify the task difficulty well above the chance. The results showed the importance of the occipital lobe (O1 and O2) coherence feature with the other brain regions. Moreover, we also study individual differences (expert vs. novice) in the classification results. The analysis revealed that the temporal lobe in experts (T4 and T3) is predominant for task difficulty classification compared with novices. 

\end{abstract}

%%%%%%%%%%%%%%%%%%%%%%%%%%%%%%%%%%%%%%%%%%%%%%%%%%%%%%%%%%%%%%%%%%%%%%%%%%%%%%%%
\section{Introduction}\label{sec:introduction}

Human-swarm interaction (HSI) has gained a lot of attention in recent years due to its many applications, including exploration, search, rescue, and surveillance \cite{kolling2013human}. The combination of mobile swarm robotics and human-in-the-loop systems generates an opportunity to leverage the best of both worlds \cite{hussein2018mixed}. However, increased task difficulty decreases the efficiency of the human operator, which in turn decreases the overall efficiency of the human-swarm team. Thus it is required to maintain an optimal level of task difficulty and intelligently adapt the task allocation between human operators and robot swarms. Hence identifying the operator's perceived task difficulty \cite{tregellas2006effect} occupies a central role in efficient HSI. 

However, the task difficulty cannot be predicted in real-time as relevant task information is not known prior to the interaction. Frequently during human-swarm interaction, the state of the swarms or swarm mission does not provide adequate information to completely quantify the nature of the task. In this direction, physiological information such as brain activity and eye movement can augment the swarm states and provide secondary information about the task and state of the operator. Thus providing a means to infer task difficulty. Specifically, by understanding the various features in the brain activity that contribute to predicting task difficulty, we allow an opportunity to predict when the operator may need assistance or is not getting enough feedback. This analysis could also potentially be used to provide insight on game balance or the training of operators \cite{ewing2016evaluation}. Moreover, due to individual differences in human brain cognition, the perceived task difficulty depends on individual differences \cite{dong2015individual}. These individual differences can be identified in a skilled marksman vs. a novice marksman. For instance, the human's visuomotor performance and efficiency have been shown to affect cortico-cortical communications in rifle shooting \cite{deeny2009electroencephalographic}. Thus, identifying different features that could be utilized for expert vs. novice operators could provide a better classification of task difficulty. In this regard, we explore the use of brain activity (measured through electroencephalography) to classify the task difficulty of a human operator interacting with a robot swarm.

To study task difficulty prediction using EEG in HSI, we have built a simulation platform that allows for human supervisory control of swarm robotics while completing missions with increasing difficulty levels. In this study, task difficulty is represented through the game difficulty levels. The simulation platform has a graphical user interface that allows the users to interact with the swarms. While the human interacts with the swarm, an electroencephalogram (EEG) is used to record the human's brain activity. We consider two approaches to classify the task difficulty level using EEG: 1) Features-based approach: Here, a set of well-known features (e.g., coherence values) are extracted and used to classify the task difficulty using a support vector machine (SVM). 2) Deep learning approach: Here, raw EEG epochs are used as input to a convolution neural network (CNN), where the output is the task difficulty level. When human participants are involved, the individual differences in skill level affect the perceived task difficulty. Hence, we consider individual differences by categorizing the participants as experts or novices and separately classifying task difficulty in each category. The results show that both feature-based SVM (71.60\%) and CNN (83.80\%) approaches can classify the task difficulty well above the chance, with CNN performing better than the SVM classifier. In terms of experts and novices, the classification accuracy is higher among experts when compared to novices. 

%%%%%%%%%%%%%%%%%%%%%%%%%%%%%%%%%%%%%%%%%%%%%%%%%%%%%%%%%%%%%%%%%%%%%%%%%%%%%%%%
\section{Method and Materials}
%%ET: In this section, we briefly discuss the simulation environment, design of experiments, human-subject study and conclude with a discussion on classification procedure.

\deleted{To explore human-swarm interaction, a gaming simulation platform that allows for synchronous physiological data streaming was developed. The simulation platform comprised three separate modules, the simulation environment, the graphical user interface, and the physiological data streaming. By utilizing these modules, the human executes a search and extract supervisory control swarm mission while their brain activity and eye movement are recorded. The mission incorporates different levels of difficulty that challenge the user to overcome different complexities as they search the environment. A non-invasive wireless EEG headset and an eye-tracking system were used to measure the physiological information during the game-play. The information of the human subject study and data collection are provided in the following sections.}

%%%%%%%%%%%%%%%%%%%%%%%%%%%%%%%%%%%%%%%%%%%%%%%%%%%%%%%%%%%%%%%%%%%%%%%%%%%%%%%%
\subsection{Human Subject Study Framework}

\begin{figure}[ht]
    \centering
    \includegraphics[width=0.75\linewidth]{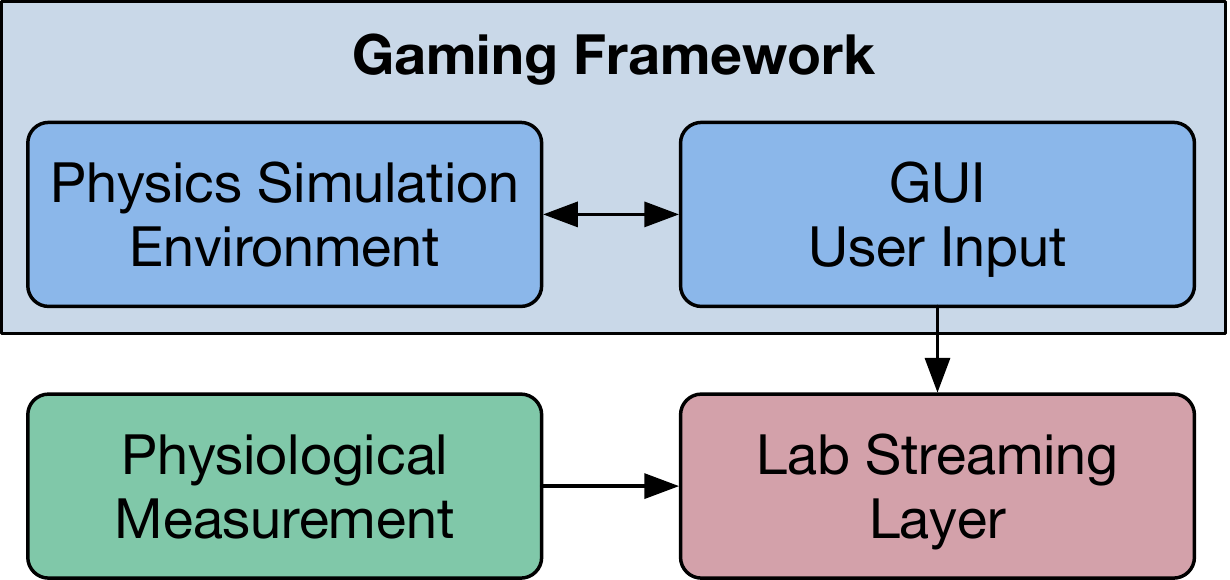}
    \caption{The framework consists of three components: Gaming environment, physiological measurement, and lab streaming layer for time synchronization.}
    \label{fig:framework}
    \vspace{-10pt}
\end{figure}

The framework for the human-swarm interaction simulation platform consisted of three separate modules (Fig. \ref{fig:framework}). The three modules are the gaming framework, the physiological data recording, and the lab streaming layer. These modules exchange information and offer different functionalities to create synchronous data recording during a human supervisory control swarm mission. These three modules run parallel using an open-source Ray framework \cite{moritz2018ray}. 

The gaming framework includes two separate components that work in parallel throughout the game. The first component is the simulation environment which was developed using the open-source library pybullet \cite{coumans2019}. The simulation environment is run through a real-time 3D physics engine. The second component is the graphical user interface that displays the information from the simulation environment. The interactive GUI presents all information needed from the simulation for the user to complete the mission.

Comprised in the simulation environment is the rendered map, which incorporates roads and buildings. In the simulation environment, the robotic swarms can execute three primitives with a high level of autonomy: formation control, path planning, and task allocation. The user is only required to input the UxV target location. In the environment, the human is supervising three UGV and three UAV platoons. More information on the design of the simulation environment can be found in \cite{manjunatha2020using}. This simulation environment was also used for learning swarm tactics over complex environments in \cite{MRS_swarm}. The lab streaming layer module is the terminal location for all data being recorded. This data includes all the environment, EEG, game, and user data. Every instance of information is inputted into the lab streaming layer, labeled with a time stamp, and stored for post-processing. 

\deleted{The simulation environment consists of three sub-modules that allow for information processing and data exchange. The three sub-modules are the state manager, the primitive manager, and the action manager. The action manager controls the execution of primitives of each platoon. The primitive manager controls which primitive the platoons are executing, and the state manager keeps a record of all the UAV, UGV, and environmental information.}

\begin{figure}
    \centering
    \includegraphics[width=0.85\linewidth]{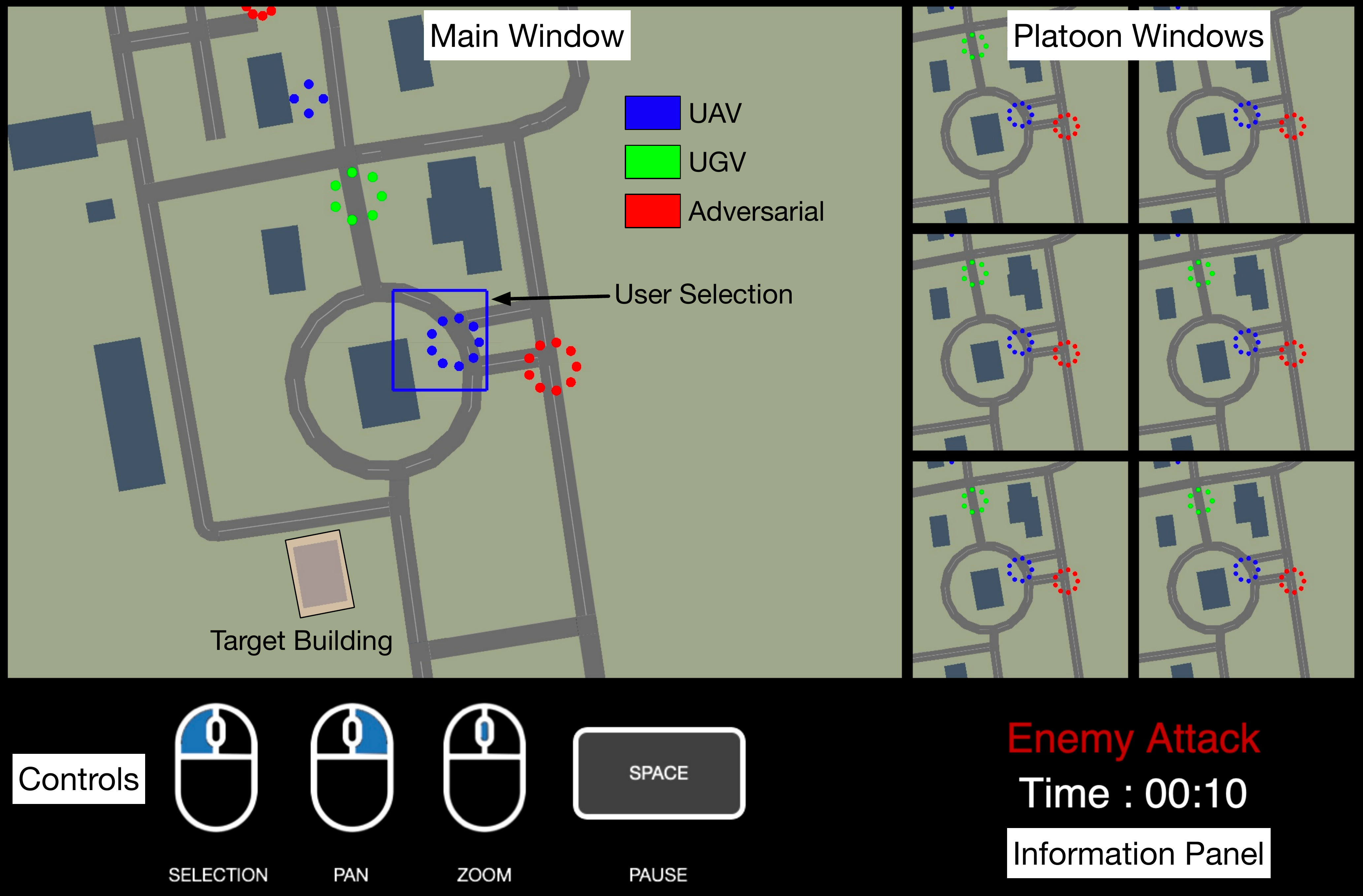}
    \caption{The GUI presents all information to the user needed to complete the mission. It allows for the control of UxV platoons through the main window. Each platoon has a panel window. The user feedback panel provides information about enemy encounters and the remaining time in the game.}
    \label{fig:gui}
\end{figure}

%%%%%%%%%%%%%%%%%%%%%%%%%%%%%%%%%%%%%%%%%%%%%%%%%%%%%%%%%%%%%%%%%%%%%%%%%%%%%%%%
\subsection{Graphical User Interface}
The simulation environment was portrayed as a 2D interactive graphical user interface using the open-source pygame library. The design of the GUI was based on popular real-time strategy games such as starcraft. The GUI consists of a partially observable environment map, platoon panels to display an up-close overview of the platoons, a user feedback panel that notifies the user of game information, and a user controls panel (Fig. \ref{fig:gui}). The GUI allows for user control allowing complete freedom to allocate the platoons anywhere on the map. More information on GUI components can be found in \cite{manjunatha2020using}.

\deleted{The GUI is designed effectively to provide easy controls for the human and provide markers for post-analysis. As an example, the user may pause the game whenever they would like to move a platoon. This allows the user to make a clear pre-meditated decision and provides a marker for when the user thought a tactical decision needed to be made in the environment. A repeated sequence of pausing, dragging a box over a platoon to select it, and picking a target location is performed to move the UAVs and UGVs to find the target.} 

%%%%%%%%%%%%%%%%%%%%%%%%%%%%%%%%%%%%%%%%%%%%%%%%%%%%%%%%%%%%%%%%%%%%%%%%%%%%%%%%
\subsection{Physiological Measurement}  

Participant's brain activity is recorded through a non-invasive B-Alert X24 EEG headset from Advanced Brain Monitoring\textsuperscript{\textcopyright} at 20 channels and 2 reference electrodes with a sampling rate of 256 Hz. The sensor locations are O1,\,O2,\,P4,\,POz,\,P3,\,Pz,\,Cz,\,C3,\,C4,\,Fz,\,F3,\,F4,\,T6,\,T4,\,F8,\,Fp1, Fp2,\,F7,\,T5,\,and T3 (Fig. 3).
  
%%%%%%%%%%%%%%%%%%%%%%%%%%%%%%%%%%%%%%%%%%%%%%%%%%%%%%%%%%%%%%%%%%%%%%%%%%%%%%%%
% \subsection{Participants}   

% 20 expert gamers were recruited from the University at Buffalo. All participants underwent a skill level screening before participating to show proficiency in real-time strategic computer gaming. Only participants that showed above-average skills were chosen for the study. Before the study, the participants had to practice and successfully complete two missions. 

\begin{figure}[b]
\includegraphics[height=1.85in, width=\linewidth]{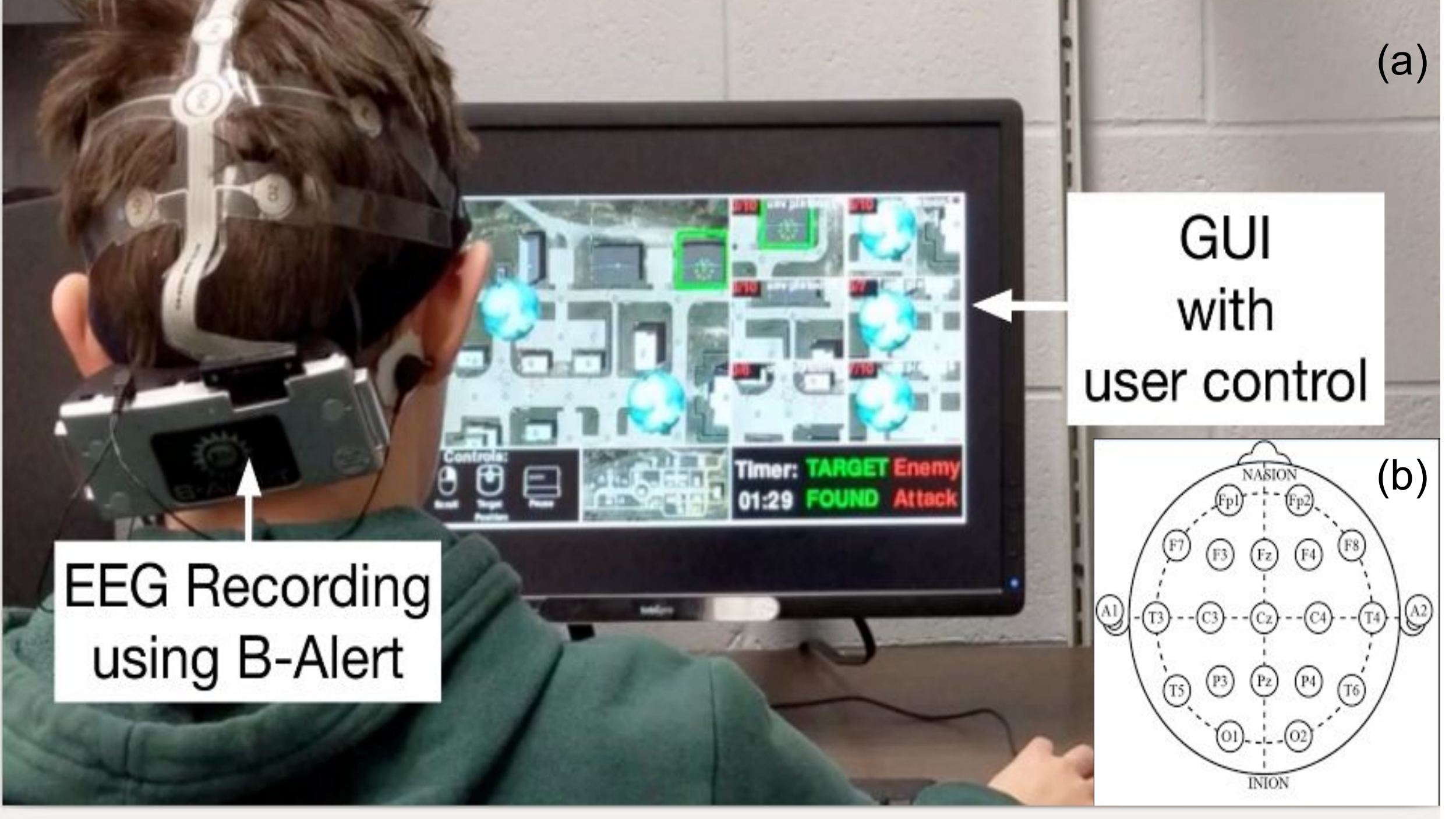}
  \caption{(a) Experimental setup for monitoring the human physiological states while supervising a group of UAVs and UGVs in target search task. (b) International EEG 10-20 electrode placements.}
\label{fig:setup}
\vspace{-10pt}
\end{figure}

%%%%%%%%%%%%%%%%%%%%%%%%%%%%%%%%%%%%%%%%%%%%%%%%%%%%%%%%%%%%%%%%%%%%%%%%%%%%%%%%
\subsection{Experiments}
Students who had experience playing real-time strategy games were recruited from the University at Buffalo. The international review board approved the human subject study (IRB \# \added{STUDY00003659}) before experimentation. All participants had normal or correct vision. The participants were asked to play the game before to show proficiency in controls. After gaining adequate practice time, the participants underwent individual difference assessments.

\deleted{When working with human subjects, individual differences are inevitable. The participant's individual differences were measured by completing two tasks that test the human's ability to multi-task and visually search an environment. The two tests being conducted are a multi-object tracking task and a visual searching test. The multi-object tracking test requires the user to keep track of multiple identified markers in a changing environment. The visual search task requires the user to identify an object in a chaotic environment quickly.}

After taking the individual difference and EEG baselines, the participants played three randomized games of different difficulties. The game's difficulties vary depending on the different complexities in the environment. The easiest task included a fully observable environment with no adversarial teams. The second level included a static adversarial team that blocked the paths and target buildings. Finally, the hardest level included a dynamic adversarial team whose position changed over time. The adversarial teams were initially hidden and made visible when the user's platoon was within the predetermined range in both static and dynamic levels. If the user's platoons and adversarial teams collided, a battle would result in the loss of platoons in a stochastic manner. If a platoon is defeated (zero members), it will disappear from the map, and the user may no longer use it.

\begin{figure}[ht]
    \includegraphics[width=\linewidth]{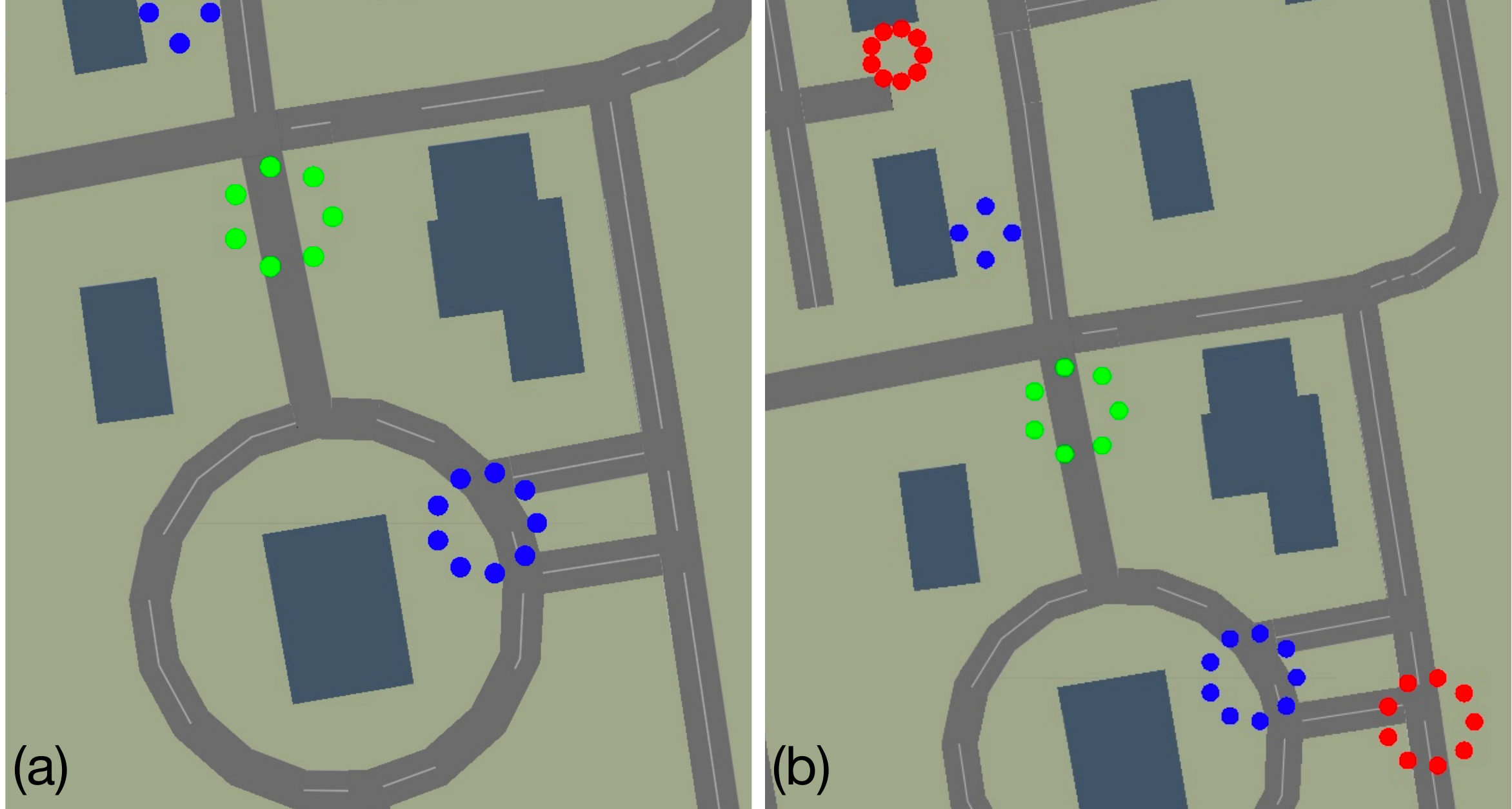}
    \caption{This figure shows the game being played both with complexities and without complexities. In a) you can see the human has no adversarial units where as in b) adversarial units (red) are blocking the paths.}
    \label{fig:Complexities}
    \vspace{-10pt}
\end{figure}

%%%%%%%%%%%%%%%%%%%%%%%%%%%%%%%%%%%%%%%%%%%%%%%%%%%%%%%%%%%%%%%%%%%%%%%%%%%%%%%%
\subsection{Individual difference}
\label{sec:individual-difference}
When working with human participants, individual differences are inevitable. Furthermore, individual differences affect the cognitive ability of humans and their perception of task difficulty. Thus it is desirable to capture this information and use it in analysis. We have used visual search (VS) and multi-object traction (MOT) to capture the individual differences \cite{memar2018physiological}. These differences allow us to categorize the participants as experts and novices. 

\begin{figure}[t]
    \includegraphics[width=\linewidth]{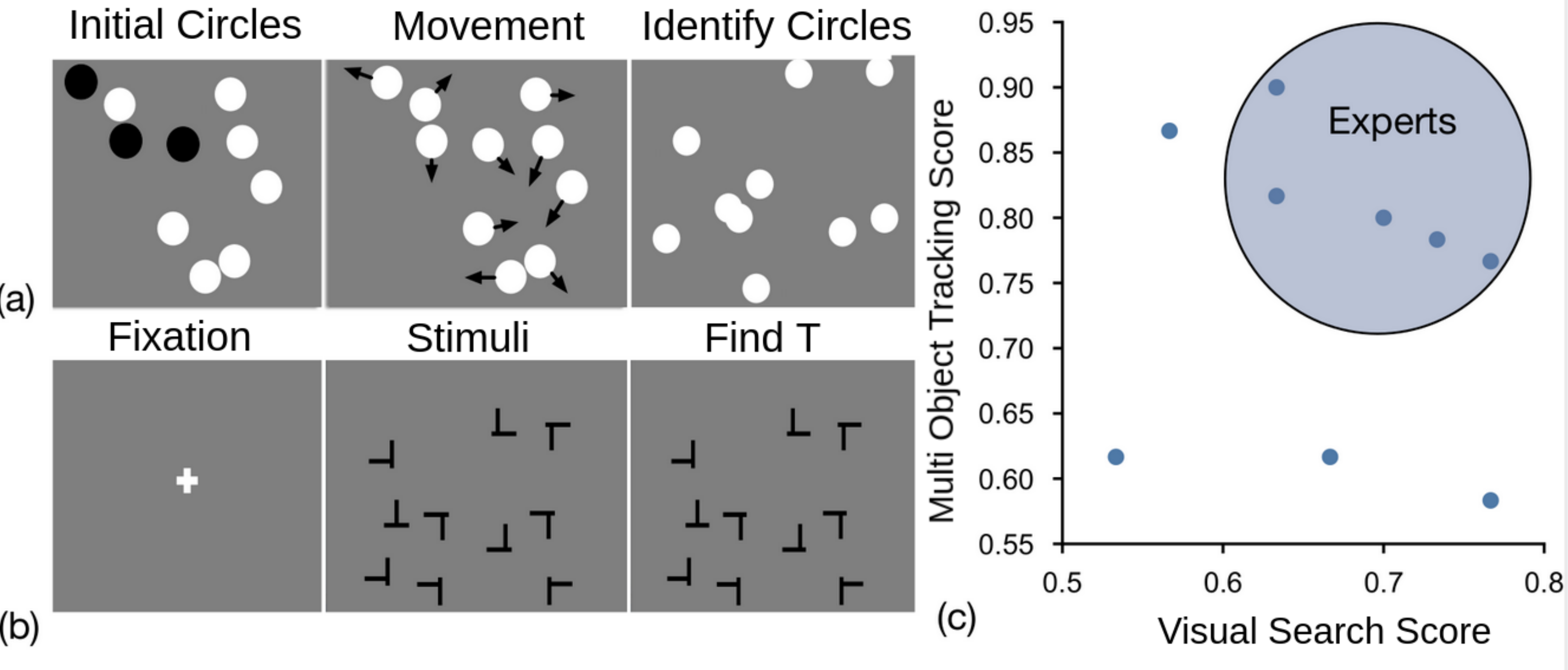}
    \caption{(a) Multi-object tracking experiment. Target objects are highlighted briefly and the participants must track them during random movement and identify them when they stop. (b) Visual search experiment. Participants are required to identify the direction of the target shape (`T') among the detractors by pressing the arrow key corresponding to the direction. (c) Identifying experts and novices using MOT and VS scores.}
    \label{fig:expert_novice}
    \vspace{-15pt}
\end{figure}

VS and MOT are EEG baseline procedures that assess individual skill levels and categorize the participant as an expert or a novice. In MOT, a participant should track a set of pre-marked circles. These circles are then moved around a space, and the user must select the marked circle at the end of the trial (Fig. \ref{fig:expert_novice}(a)). VS is tested by flashing a screen with multiple shifted `T's and asking the user to identify the direction of the correct T within a 2-second window (Fig. \ref{fig:expert_novice}(b)). The user's scores depend on their ability to identify the circles and find the direction of the T. If the participant scored well on both tests, they are considered an expert. Out of 9 participants, 5 were labeled as experts, and the remaining were labeled as novices.

%%%%%%%%%%%%%%%%%%%%%%%%%%%%%%%%%%%%%%%%%%%%%%%%%%%%%%%%%%%%%%%%%%%%%%%%%%%%%%%%
\section{Data Analysis} 
\deleted{The data recorded during the human subject study was processed for feature extraction analysis, feature selection analysis, classification using a linear classifier, and a deep learning network. The details of the data analysis techniques are elaborated in the section below.}

The data recorded during the human subject study underwent post-processing and further analysis. The details of the data analysis techniques are elaborated in the section below.

%%%%%%%%%%%%%%%%%%%%%%%%%%%%%%%%%%%%%%%%%%%%%%%%%%%%%%%%%%%%%%%%%%%%%%%%%%%%%%%%
\subsection{Classification}
\label{sec:classification}
Classifying task difficulties is posed as a three-class classification problem. The baseline mission data with no adversarial team \replaced{is the least difficult level}{was the first least difficult}. \added{The difficulty level is increased by introducing static and dynamic adversarial teams. Before classification, the class labels were balanced. We explored two classification methods, support vector machine (SVM) classifier with EEG features and Convolutional neural network (CNN) approach with raw EEG.}% The classifier was a linear support vector machine (SVM).}

\begin{figure}[t]
    \centering
    \includegraphics[width=0.75\linewidth]{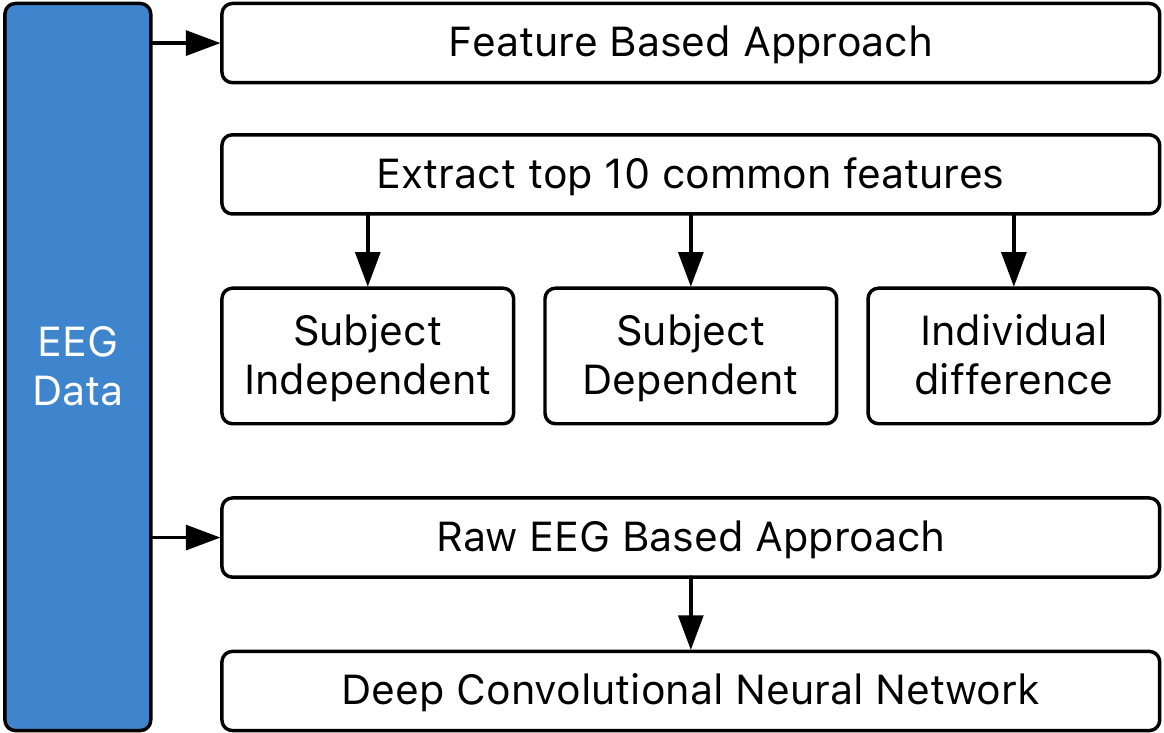}
    \caption{Data splitting scheme for classification.}
    \label{fig:data-splitting} 
    \vspace{-10pt}
\end{figure}

\added{We have considered three types of data splitting schemes. Due to the individual differences in EEG research, it is important to explore different training data sets to obtain accurate feature selection results and compare the results obtained from different training sets. 1) Subject independent: pool all the participant's data, balance, and randomly split 66\% for training and 33\% for testing. 2) Subject dependent: Out of 10 participants, training is done in 9 participants, and testing is done on the remaining participants (leave-one-out). The process is repeated until every participant is considered for testing. 3) Individual difference-based: The participants are split into experts and novices depending on the individual difference score (see Section \ref{sec:individual-difference}). Then the classification is carried separately for experts and novices. The discussed data splitting scheme is shown in Fig. \ref{fig:data-splitting}. Note that the data is split after the feature selection step, and the scheme is the same for SVM and CNN.}

\deleted{The data was split in two ways to consider subject independence and subject dependence. The data was split randomly for subject independence, so 2/3 were used for training, and 1/3 was used for testing. For subject dependence, the data was trained on 8 subjects and tested on two.}

%%%%%%%%%%%%%%%%%%%%%%%%%%%%%%%%%%%%%%%%%%%%%%%%%%%%%%%%%%%%%%%%%%%%%%%%%%%%%%%%

\subsection{Feature Extraction}
\deleted{The mission had different levels of difficulty to challenge the user. The baseline mission was easy due to the environment being fully known without any adversarial. The mission got more difficult when the dynamic adversarial teams surrounded the target buildings and roamed the map. These adversarial teams were only seen when the user had a platoon close to the adversarial team, making the map partially observable. Due to the adversarial teams, the user had to make more decisions, understand the environment, and change the paths of the platoons more often. The change in task difficulty directly affects the cognitive load of the human playing the game \cite{de2019brain}.}

The EEG signals were bandpass filtered (0.1-70 Hz) to remove the electrical noise of the headset and the slow drift caused by the electrodes moving on the scalp. EEG signals were epoched into 2-second windows, and epochs with large artifacts were dropped. Increasing the window length would increase the noise, and decreasing the length would reduce the data points available for learning. Hence, we heuristically choose a window length of 2 seconds. The eye blink and drastic body movement artifacts were also removed using independent component analysis with the Picard algorithm in MNE-Python~\cite{ablin2018faster}. 

Once a cleaned epoched EEG signal was obtained, the features were extracted. The first set of features extracted were from the b-alert live software (Advanced Brain Monitoring\textcopyright. Carlsberg, CA, USA). These features include mental workload, engagement, and distraction. Additional details on feature extraction and baselines used can be found in \cite{berka2007eeg}.

The second set of features were coherence values from the brain activity frequency bands. Coherence measures the similarity of two signals in a predefined frequency band for a given period. In this paper, we have used 6 frequency bands; delta [4-8 Hz], low alpha [8-10 Hz], high alpha [11-13 Hz], low beta [14-22 Hz], high beta [23-35 Hz] and gamma [36-44 Hz]. The coherence values (as given by equation \ref{eqn:coherence}) were calculated for each electrode pair and frequency band.

\begin{equation}
    \label{eqn:coherence}
    coh = \frac{ \lvert E[Sxy] \rvert}{ \sqrt{E[Sxx]*E[Syy]}}
\end{equation}

\noindent where $S_{xx}$, $S_{yy}$ are power spectral densities of the channels, and $S_{xy}$ is the cross power spectral density of the channel being used. \deleted{From the baseline testing features and the coherence, there were a total of 1148 features for each two-second epoch window.}
%%%%%%%%%%%%%%%%%%%%%%%%%%%%%%%%%%%%%%%%%%%%%%%%%%%%%%%%%%%%%%%%%%%%%%%%%%%%%%%%

\subsection{Feature Selection}
\deleted{The goal of this work is to explore the feature space to deduce what EEG features are significantly important to task difficulty in human swarm interaction. A principal component analysis (PCA) was done first to discover the amount of variance covered by the features. While performing PCA, it was found that the data was widespread. Two principal components contributed to 0.32 of the variance, while the other components make up the rest. When inspecting the principal components, it was seen that select features have effects that allow us to know that significant features exist.} 

\added{Since the number of features is larger, the classification algorithm can over-fit by learning noise. To avoid over-fitting, we perform feature selection. The present study recorded EEG from 20 channels using a 10-20 international system and considered six frequency bands for feature extraction. This results in 1200 coherence features. To reduce the number of features even further, we explored the existing literature on task difficulty. We found that the frontal, parietal, motor, and occipital regions, especially during visual-motor tasks, are predominantly correlated with task difficulty \cite{memar2019objective}. Thus we selected T4, T3, O1, P3, Pz, F3, Fz, F4, C4, P4, C3, Cz, and O2 electrodes to use.}

A recursive feature elimination (RFE) was performed to obtain the best set of features. The RFE takes the initial set of features used as the input for the machine learning algorithm and ranks the features. The features deemed significant stay in the data set, and the features that have small effects on the classification are removed. This is repeated until a user-defined number of features are identified. \added{RFE is stochastic in nature. As a result, different features can be selected in a different trial. To reduce the effect of randomness, we ran RFE 10 times and selected the most common features. For RFE, we have used linear SVM with RBF kernel.}

\deleted{In this paper, five features are desired, and the machine learning algorithm used is a support vector regression with a linear kernel. The features with the greatest weights are kept through every iteration. 
10 fold cross-validation was used to determine the features with the most significance.}
%%%%%%%%%%%%%%%%%%%%%%%%%%%%%%%%%%%%%%%%%%%%%%%%%%%%%%%%%%%%%%%%%%%%%%%%%%%%%%%%
\subsection{Convolutional neural network}

\begin{figure}[h]
    \centering
    \includegraphics[width=0.85\linewidth]{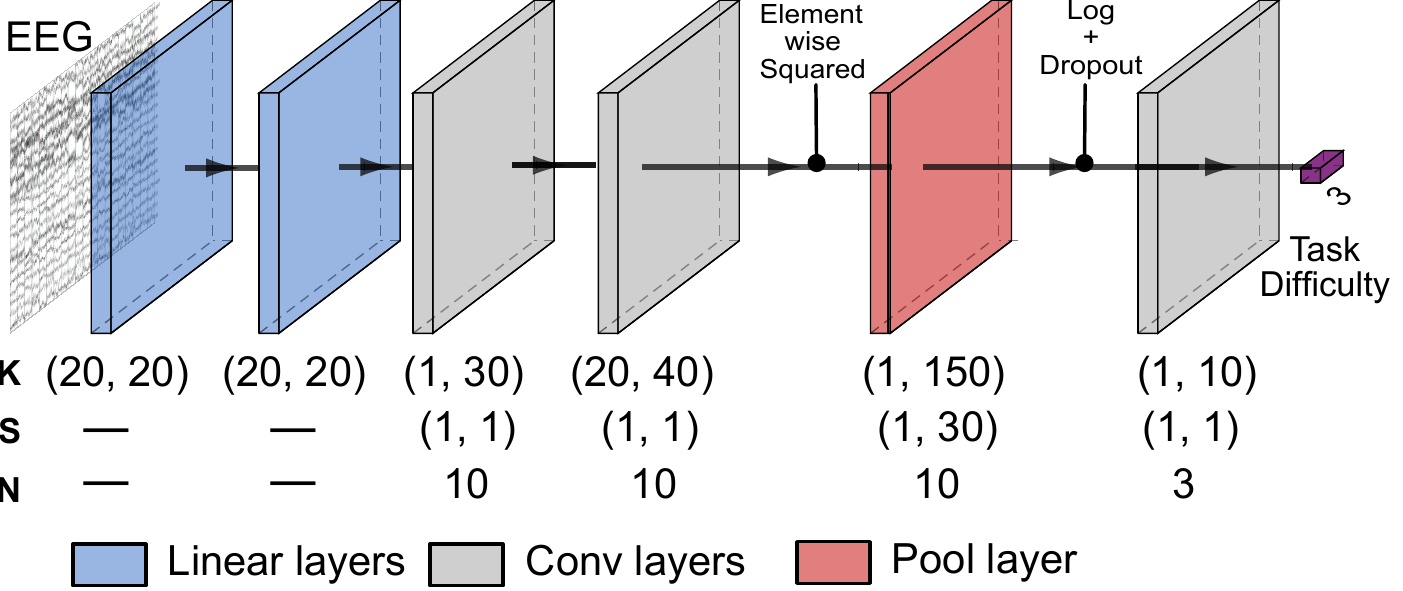}
    \caption{Architecture of the deep learning model. K: Kernel size, S: Stride, and N: Number of filters.}
    \label{fig:cnn}
\end{figure}

The architecture of the convolution neural network (CNN) used for classification, as shown in Fig. \ref{fig:cnn} is inspired by \cite{schirrmeister2017deep} and \cite{passalis2019deep}. The first two linear layer operation is given by 
\begin{equation}
 \label{eq:1inear-layers}
    x' = (x-W_{sh}\bar{x}) \odot (W_{sc}\tilde{x}_{sh})
\end{equation} 

Where $x$ is the input epoch, $\bar{x}$ is the mean value calculated across each electrode and $\tilde{x}_{sh}$ is the standard deviation of $(x-W_{sh}\bar{x})$. Note $W_{sh}$ and $W_{sc}$ are the learnt weights of the linear layers and the $\odot$ represents element-wise multiplication (Hadamard product). Note that if$W_{sh}$ is identity and $W_{sc}$ is the inverse covariance matrix (precision matrix), then the equation \ref{eq:1inear-layers} reduces to a standard z-score.
The linear layers are succeeded by two convolution operations; kernel size $1\times30$ and $20\times40$ (stride of 1). The number of filters is kept as 10 for both convolution layers. After two convolution layers, the output is element-wise squared, followed by average pooling. The kernel for average is $1\times150$, and stride is $1\times30$. After pooling, the $log$ function is applied to the output, followed by a dropout. The dropout percentage is kept at 20\% to avoid over-fitting. Finally, a convolution operation with kernel size $1\times10$ (stride=1) is applied to reduce the feature size to 3. Finally, a softmax function is used to convert the output to class probability.
%%%%%%%%%%%%%%%%%%%%%%%%%%%%%%%%%%%%%%%%%%%%%%%%%%%%%%%%%%%%%%%%%%%%%%%%%%%%%%%%
\section{Results and Discussion}
\added{In this section, we provide the classification results and discuss the importance of selected features. Note the discussion of selected features is in the context of the SVM classifier. Table \ref{tab:classification_results} shows the classification results using two different methods and different data splitting schemes (see Section \ref{sec:classification}). For subject-independent classification, deep learning performed better than the feature-based SVM classifier. However, for subject-dependent classification, SVM performance was better than the deep learning approach. Nonetheless, both classifier's performance is poor for subject-dependent classification. The main reason for poor performance can be attributed to individual differences among participants. Interestingly, when the classification is performed separately between expert and novice, the classification rate is higher for experts than novices. This trend holds both for the SVM and CNN approaches. A probable explanation might be that experts make more efficient use of neural resources leading to different brain dynamics when compared with novices \cite{dong2015individual}.} 
\deleted{The classification results from the deep learning and raw EEG yield better results for a more granular three-class classification. Deep learning performs better than the SVM classifier for three-class classification. All classifications yield a result 30 percent above chance.}

\setlength{\tabcolsep}{5pt}
\begin{table}[htbp]
  \centering
  \caption{Classification results}
    \begin{tabular}{ccccc}
    \hline
    \multicolumn{1}{P{4.585em}}{\textbf{Classification \newline{}Method}} &
    \multicolumn{1}{P{4.585em}}{\textbf{Subject \newline{}Independent}} &
    \multicolumn{1}{P{4.585em}}{\textbf{Subject \newline{}Dependent}} &
    \multicolumn{1}{P{4.585em}}{\textbf{Expert}} & 
    \multicolumn{1}{P{4.585em}}{\textbf{Novice}}\\
    \hline
    CNN & 83.80$\pm$1.42 & 42.19$\pm$2.88 & 90.11$\pm$4.87 &  86.13$\pm$5.63\\
    SVM & 71.60$\pm$2.49 & 50.1$\pm$1.23 & 77.98$\pm$1.65 & 72.14$\pm$1.05\\
    \hline
    \end{tabular}%
  \label{tab:classification_results}%
\end{table}%

The significant features extracted from the feature selection are listed in Table \ref{tab:selected_features}. Among the selected features, the Pz-O2 and O1-P4 coherence values show the inner connectivity in the parietal and occipital lobes of the brain. The F3-C3 and F3-O2 coherence values portray the connectivity between the prefrontal cortex with the brain's motor cortex and occipital lobe. These features signify which sections of the brain interact differently during different task difficulties. 

\begin{table}[htbp]
    \centering
    \caption{Significant Features}
    \begin{tabular}{c|c}
     \hline
     \textbf{Coherence Electrodes} & \textbf{Frequency Band}\\
     \hline
     Pz-O2 & 8-10 Hz  (Low Alpha)\\
     F3-C3 & 14-22 and 23-35 Hz (Low and high beta)\\
     F3-O2 & 14-22 and 23-35 Hz (Low and high beta)\\
     O1-P4 & 14-22 and 35-44 Hz (Low beta and gamma)\\
     \hline
    \end{tabular}
    \label{tab:selected_features}%
\end{table}%

\begin{table}[htbp]
    \centering
    \caption{Expert vs Novice Significant Features}
    \begin{tabular}{c|c}
     \hline
     \textbf{Expert Features} & \textbf{Novice Features}\\
     \hline
     T4-O2 (High beta 23-35 Hz) & O1-C4 (Low Beta 14-22 Hz)\\
     T3-O1 (High beta 23-35 Hz) & Cz-O2 (Low Beta 14-22 Hz)\\
     Fz-F4 (Gamma 36-44 Hz) & Pz-C4 (Low Beta 14-22 Hz)\\
     \hline
    \end{tabular}
    \label{tab:selected_features_expert_novice}%
\end{table}%

\begin{figure}[t]
    \includegraphics[height=85pt, width=\linewidth]{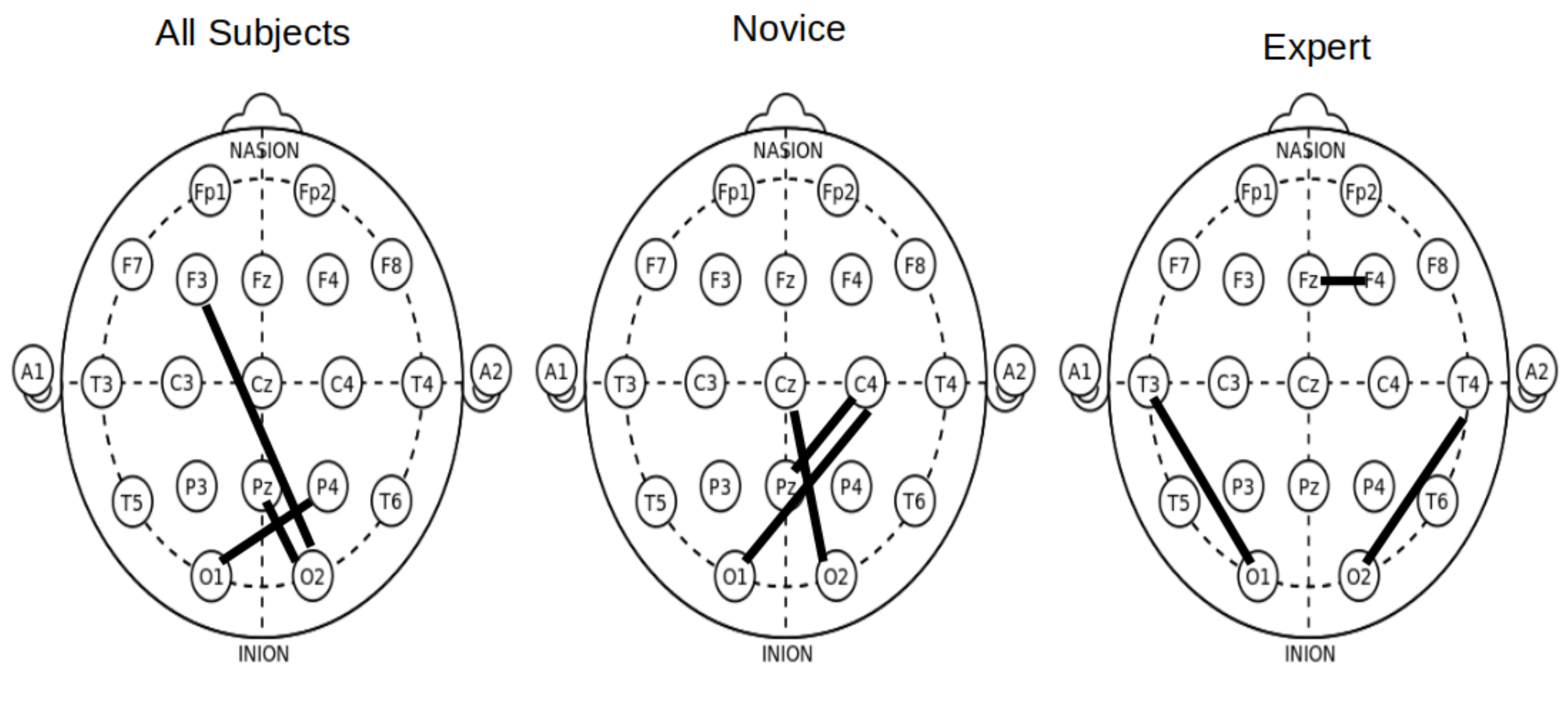}
    \caption{The electrodes that have significant coherence are displayed on the EEG 10-20 system.}
    \label{fig:brain_con} 
    \vspace{-10pt}
\end{figure}

The expert features show a strong correlation between the temporal lobe and the occipital lobe. These are unique features that did not show up in the previous feature selection. The novice features all contained electrodes in the motor cortex. Fig. \ref{fig:brain_con} shows the visual representation of the significant features on the EEG 10-20 electrode locations. As you can see, all participant's features connect the prefrontal cortex and the occipital lobe. However, the expert's connectivity features are unique because they connect the occipital lobe to the temporal side lobes. 

\deleted{The deep learning framework and a linear classifier were able to predict task difficulty at a high level to show that the task difficulty impacts the brain activity during human swarm interaction. Although the accuracy of the SVM classifier could have been much higher by adding more coherence features, the feature space would have been too saturated to find significant features. The SVM also predicted the above chance for the subject-dependent data splitting and selected the same significant features as the independent data split.} 

\deleted{The main contribution of this work is deducing the features found significant for classifying task difficulty in human swarm interaction. The significant features chosen support previous research on task difficulty.} \added{In terms of most selected features,} the Pz-O2 coherence is representative of the occipital and parietal lobe connectivity. In previous research, the low alpha band has reports of synchronization and variance when performing different tasks. \cite{korobeinikova2012spectral}, \cite{gevins2003neurophysiological}. It was also found that the Pz electrode spectral density drastically changes due to task load in human-computer interaction with an event-related potential \cite{smith2001monitoring}. The beta frequency band's F3-C3 and F3-O2 coherence values represent the prefrontal cortex connectivity with the motor cortex and the occipital lobe. The occipital lobe activation indicates the visual demand of humans, and the frontal lobe signifies engagement and data processing \cite{rietschel2012cerebral}. In this human swarm interaction mission, both the visual demand and the engagement would increase due to the added adversarial complexities. \added{On similar lines}, the O1-P4 electrode \deleted{does not hold as much value as the other features but}can also be accredited to the increase in visual demand. 

The selected features between experts and novices are significantly different. The expert features are unique due to the involvement of the temporal lobe. Previous research shows that electrodes Fz, T4, and T3 power spectral densities strongly correlate with psychomotor efficiency. This has been shown both in rifle shooting by experts and novices and through a physical human-robot interaction task difficulty analysis \cite{memar2018eeg, deeny2009electroencephalographic}. Therefore, these three electrodes are included in the expert features but are lacked in the novice features. Although the Fz, T4, and T3 coherence features are not selected, the connection with the occipital lobe could be from the visual intensity of the human swarm interaction task.
%%%%%%%%%%%%%%%%%%%%%%%%%%%%%%%%%%%%%%%%%%%%%%%%%%%%%%%%%%%%%%%%%%%%%%%%%%%%%%%%
\section{Conclusion}
This paper presents an analysis of brain activity during a human swarm interaction study \added{to quantify task-difficulty in a target search mission}. \deleted{A new framework for researching human swarm interaction was explained, and the data was used to analyze task difficulty from a search and extract supervisory control mission.} The participants played three separate games to find the target location with increasing task difficulty due to adversarial complexities in a partially observable map. \added{While the participants provided supervisory commands, their brain activity was measured, which is used to quantify the task difficulty level.}

\added{Task difficulty quantification was posed as a three class classification problem using EEG data as the input. The three-class problem corresponds to the three different levels of difficulty in the mission: no adversarial team, static adversarial team, and dynamic adversarial team. We explored two classification methods: support vector machine (SVM) classifier with EEG features and deep learning approach with raw EEG. The EEG features consisted of coherence values extracted from 2-second epochs from T4, T3, O1, P3, Pz, F3, Fz, F4, C4, P4, C3, Cz, and O2 sites and in the alpha, beta, and gamma frequency bands. For the deep learning approach, the raw epoch was directly used as an input. Irrespective of the classification method, the EEG signals were filtered and cleaned with ICA. Both feature-based and deep learning approaches were able to classify task difficulty well above the chance (33\%). Most of the prominent features in the classification showed a connection between the prefrontal cortex with the motor cortex and occipital lobes.}

Due to the individual differences \replaced{in the skill level}{caused by EEG}, the participants were separated into experts and novices based on \deleted{their baseline results in} the multi-object tracking and visual search tasks scores. \added{Among the experts, coherence values at the temporal lobe (T3 and T4) were most significant, indicating that the experts have more psychomotor efficiency than novices. However, most of the selected coherence features involved the occipital lobe regardless of expert or novice.} In future work, we wish to explore the relationship between task difficulty classification and the cognitive workload of the participants. We also hope to explore more complex features including the human brain connectivity graphs. 

\deleted{The feature selection was performed again, and it was found that the expert's temporal lobe electrodes are significant. This indicates that the experts have psychomotor efficiency. It was also seen through all the features that human swarm interaction is incredibly visually demanding based on the task. Most of the features that were selected involved the occipital lobe regardless of expertise.}

\deleted{It should be noted that there is a high amount of variance in the data that could lead to low classification results and a large number of features that could be considered significant. Consequently, the number of data used and features used needed to be less to account for this.}
%%%%%%%%%%%%%%%%%%%%%%%%%%%%%%%%%%%%%%%%%%%%%%%%%%%%%%%%%%%%%%%%%%%%%%%%%%%%%%%%

\printbibliography

\end{document}